
----------
X-Sun-Data-Type: default
X-Sun-Data-Description: default
X-Sun-Data-Name: sdymr.tex
X-Sun-Content-Lines: 720

\magnification=\magstep1
\baselineskip=14truept
\overfullrule-0pt

\hsize=6truein
\vsize=8.5truein

\long\def\Theorem#1{{\noindent \bf Theorem} {\it #1 }}

\def\gsl_2{{\underline {s\ell}}_2}

\def\reals{{{\rm I} \kern -.15em {\rm R}}}
\def\C{{\rm \kern.24em \vrule width.02em height1.4ex depth-.05ex
\kern-.26em C}}
\def\ints{{{\rm Z} \kern -.35em {\rm Z}}}

\def\nvarepsilon{{{\varepsilon}\kern -.47em{\backslash}}}
\def\nvarep{{{\varepsilon}\kern - .40em{|}}}

\def\ovy{{\overline y}}
\def\ovz{{\overline z}}
\def\underg{{\underline {g}}}
\def\underp{{\underline {p}}}
\def\underq{{\underline {q}}}
\def\undert{{\underline {t}}}

\font\caligraphic=cmsy10 at 10pt
\def\cal#1{{\caligraphic #1}}

\def\DD{{\hbox{\cal D}}}
\def\EE{{\hbox{\cal E}}}

\def\GG{{\hbox{\cal G}}}
\def\HH{{\hbox{\cal H}}}

\def\MM{{\hbox{\cal M}}}



\centerline{\bf
On Soliton Content of}
\centerline{\bf Self Dual Yang-Mills Equations}

\vskip.3truein

\centerline{\bf J. Szmigielski}
\centerline{\it Department of Mathematics and Statistics}
\centerline{\it University of Saskatchewan}
\centerline{\it Saskatoon, SK\qquad S7N OWO, Canada}

\bigskip

\baselineskip=12truept

\centerline{{\bf ABSTRACT}}

{\narrower\smallskip Exploiting the formulation of the Self Dual Yang-Mills
equations as a Riemann-Hilbert factorization problem, we present a theory
of
pulling back soliton hierarchies to the Self Dual Yang-Mills equations.
We show that for each map $
\C^4 \to \C^{\infty } $ satisfying a simple system of linear equations
formulated
below one can pull back the (generalized) Drinfeld-Sokolov hierarchies to the
Self Dual Yang-Mills equations.  This indicates that there is a class of
solutions to the Self Dual Yang-Mills equations  which can be constructed
using the soliton techniques like the $\tau$ function method.
In particular this class contains the solutions obtained via
the symmetry reductions of
the
Self Dual Yang-Mills equations.  It also contains  genuine 4 dimensional
solutions . The method can be used to study the symmetry reductions
and as an example of that  we get an
equation exibiting breaking solitons,
formulated by O. Bogoyavlenskii,  as one of the
$2 +
1 $ dimensional reductions of the Self Dual Yang-Mills equations.
\smallskip}

\bigskip

\baselineskip=18truept
\noindent
{\bf 1. Introduction.}
\medskip

The Self Dual Yang-Mills (SDYM) equations are an integrable system
admitting
the zero-curvature representation involving a spectral parameter.  Denoting
the
spectral parameter by $\lambda $, where for simplicity we take $ \lambda
\in S^1
$, we can write the zero curvature condition as
$$
\left[\left( {\partial\over\partial\,\ovy} + \lambda
{\partial\over\partial z}\right) - A_\ovy - \lambda A_z ,\
\left(-{\partial\over
\partial\,\ovz} + \lambda {\partial\over \partial y}\right) + A_ \ovz -
\lambda A_y\
\right] = 0\ .  \leqno (1)
$$
In the physical coordinates $x = (x_1, x_2, x_3, x_4) \in\reals ^4$,
$$
y = x_1 + ix_2,\quad z = x_3 - ix_4   \leqno (2)
$$
and $\ovy,\ \ovz$ are complex conjugates.

It was R. Ward [1], who interpreted (1) as a differential consequence of an
assumption of holomorphic triviality over projective lines.  The most
succinct
formulation is via a Riemann-Hilbert factorization
problem.  Let $ w_1 = y + \lambda \ovz ,\hfil\break
 w_2 = -\ovz + \lambda y,\ \partial _1 =
{\partial\over \partial\ovy} + \lambda {\partial\over \partial z},\
\partial _2
= -{\partial\over \partial\,\ovz} + \lambda {\partial\over \partial y} $.
Then,
Ward's triviality condition reads:
$$
g (w_1, w_2, \lambda ) = \phi^{-1}_{-} (x,\lambda ) \phi _{+} (x,\lambda )
\ ,\ g \in SL(n,\C),\ \partial _1 g = \partial _2 g = 0 \quad\hbox
{and}\quad
\phi _{+}(\phi _{-}) \leqno (3) $$
are matrices in $ SL(n,\C) $, holomorphic with respect to $ \lambda $
inside (outside) the unit circle.  For the sake of our discussion we need a
generalization of equation (3).  Consider $ \GG $ and $ \HH $, both belonging
to $
SL(n,\C) $ and both functions of $ \lambda $ and $ x $, satisfying
$$
\partial _1 \GG = \Omega ^{-}_1 \GG,\ \partial _1 \HH = \Omega ^{+}_1 \HH, \
\partial _2 \GG =
\Omega ^{-}_2 \GG,\ \partial _2 \HH = \Omega ^{+}_2 \HH\ ,  \leqno (4)
$$
where
$$
\Omega ^{+}_i (x,\lambda ) = \sum ^{\infty }_{k=0} \omega
^{+}_{ik}(x)\lambda
^{k},\  \Omega ^{-}_i (x,\lambda ) = \sum ^1_{k=-\infty }\omega ^{-}_{ik}
(x)\lambda ^k ,\quad i = 1,2\ .
$$
Then, the existence (and differentiability) of the Riemann-Hilbert
factorization
$$
\GG \HH^{-1} = \phi ^{-1}_{-} \phi _{+}  \leqno (5)
$$
implies that the analog of the soliton Baker-Akhiezer function $ \psi =
\phi _{-} \GG =
\phi _{+}\HH $ satisfies
$$
\partial _1 \psi = (A_\ovy + \lambda A_z) \psi  \leqno (6a)
$$
$$
\partial _2 \psi = (-A_{\ovz} + \lambda A_z)\psi  \leqno (6b)
$$
where $ (A_y, A_\ovy , A_z, A_\ovz ) $ is a self-dual connection.  This
statement is a simple
consequence of equations (4) and (5) and Liouville theorem.
\bigskip

\noindent
{\bf 2. Generalized Drinfeld-Sokolov hierarchies.}
\medskip

A generalization of Drinfeld-Sokolov flows is presented fully in [2]. The
canonical reference for the Drinfeld-Sokolov systems is [3]. Among
the facts we
shall need below are the following.  Consider the Kac-Moody algebra
$A^{(1)}_{n -
1}$([4]).  The canonical generators are
$e_0 = \pmatrix { 0 & \ldots & 0\cr
        0 && \vdots\cr
        \vdots && \vdots\cr
        \lambda & \ldots & 0 }$ and \hfil\break
$e_k = \pmatrix{ 0 & 0  & & &\cr
                  & \ddots & \ddots &  &\cr
                  & & \ddots & 1 &\cr
                  & & & \ddots & \ddots \cr
                  & & & & 0 & 0}$ \quad
$ 1 \le k \le n - 1$, \quad $f_0(\lambda) = e^t_0 \left({1 \over
\lambda}\right)$,\quad
$f_m = e^t_m$.  The cyclic element $J_\lambda = \sum\limits^{n-1}_{k=0}
e_k$ has the
property $J_\lambda^{kn} = \lambda^k$.

Let $\Delta = \Delta _- \dot\cup \Delta_+$ be the set of all roots of
$g =
A^{(1)}_{n-1}$, and \hfil\break
$\pi = \{\alpha_0, \dots, \alpha_{n-1}\}$ be the set of simple
roots.   $\underg $ admits the decomposition:
$$
\underg = \underg_- \oplus \underg_0 \oplus \underg_+\ ,
$$
where
$$
\underg_{\pm} = \bigoplus\limits_{\alpha \in \Delta_\pm} \underg _ \alpha\
{}.
$$
Recall that a subalgebra $\underp$ of $\underg$ is called a
\underbar{parabolic} \underbar{subalgebra} if
$$
\underg_0 \oplus \underg_+ \subseteq \underp\ .
$$
For any parabolic subalgebra $\underp$ there is a subset $\tilde \pi$ of
$\pi$, such that if
$$
\tilde\Delta_+ = \left\{ \sum k_i\alpha_i \in \Delta_+ : \alpha_i \in
\tilde
\pi\right\}\ ,
$$
then $\underp = \left( \bigoplus\limits_{\alpha \in \tilde\Delta_+}
\underg_{-\alpha}\right)
\oplus \underg _0 \oplus \underg _+$.
If, consequently, one defines $\underq = \bigoplus\limits_{\alpha \in
\Delta+
\backslash\tilde\Delta_+}\underg _\alpha$, then
$$
\underg = \underq \oplus \underp \ .\leqno (7)
$$
In our context, we assume that $\tilde \pi $ \underbar {does} \underbar
{not}
contain $\alpha_0$, that is $f_0$ does not belong to $\underp$.  Thus
$\underp$
differs from $\underg_0 \oplus \underg _+$ by some lower triangular
matrices
independent of $\lambda$.  In fact this is just a lower triangular piece of
a parabolic subalgebra of $sl(n, \C)$.  We fix now $\underp$ and $\underq$ and
form the
corresponding connected groups $G_p$ and $G_q$, both subgroups of $G=\{$
space
of maps: $S^1 \to SL(n, \C)\}$.  The generalized Drinfeld-Sokolov flows can
be
defined through the factorization problem
$$
e^{\sum\limits_{k\in\ints}t_kJ^k_\lambda}
ge^{-\sum\limits_{k\in\ints}t_kJ^k_\lambda}
 = \phi^{-1}_- (\undert) \phi_+ (\undert)\ , \leqno (8)
$$
where $\phi_- \in G_q,\quad \phi_+ \in G_p,\quad g \in G$\ .  For all
practical
purposes we assume that all but a finite number of $t_k{\rm 's}$ vanish.  The
equations of (generalized) Drinfeld-Sokolov type arise as the compatibility
conditions of
$$
{\partial \over {\partial t_k}}\psi = U_k \psi\ , \qquad k \in \ints\ ,
\leqno (9)
$$
where $\psi$ is the Baker-Akhiezer function
$$
\psi(\undert) = \phi_- (\undert) e^{\sum\limits_{k\in\ints}t_kJ^k_\lambda}
\leqno(10a)
$$
and $U_k $ can be computed from (8) via
$$
U_k = ( \phi_-J^k_\lambda \phi_-^{-1} )_p + ( \phi_+J^k_\lambda \phi_+^{-1} )_
q  \leqno(10b)
$$
where the subscripts p and q denote the projections on $\underp$ and $\underq$.
The theory depends in an essential way on the sign of k.  For  $k>0$ we have a
well known statement [3] that $U_k$ is a differential polynomial
(with respect to the first flow $t_1=x$) in the entries of $U_1 $.   As a
result the corresponding compatibility condition is an evolution equation.
This is not so for $k<0$.   To illustrate that we quote a few results from [2].
For $n = 2,\ \tilde \pi = \{\alpha _1\}, \ k = 1, 3$

$$
U_1 = J_\lambda +q,$$
$$
q=\pmatrix {u & 0\cr Du-u^2 & -u },
$$
and the compatibility of the first and the third flow gives
the
potential $KdV$ equation $u_t = {1 \over 4} u_{xxx} + {3 \over 2} (u_x)^2$.
On the other hand
for $k = 1, -1$ one obtains the stationary Bogoyavlenskii equation [5], [2],
$u_{xxxt}
=
4(u_{xx} + u_{xx}\cdot u_t + 2u_x u_{xt})$.  When $\tilde \pi = \emptyset$, one
gets
the modified $KdV$ hierarchy.  Including the "negative" flows in this case is
well known to yield the Toda equations.
\bigskip\bigskip\bigskip
\noindent
{\bf 3. Solving  SDYM using soliton hierarchies.}
\medskip

Now suppose $t_k{\rm 's}$ are functions of the variables appearing in the
SDYM
equations, namely $(y, \ovy, z,\ovz)$.  The question we want to address
is
that of admissible choices of a functional dependence of generalized
Drinfeld-Sokolov flows for which the factorization problem (8) goes into
the
factorization problem (5).
In other words we are looking for a map
$\phi : \C ^4 \to \C ^{\infty},(y,\ovy,z,\ovz)\to \undert =
\{t_k\}_{k\in\ints}$
which pulls back (9) to (6a) and (6b).
To this end the functions

$$
\GG = e^{\sum\limits_{k\in\ints}t_k J^k_\lambda }\quad and \quad
\HH=e^{\sum\limits_{k\in\ints}t_k J^k_\lambda }g \ , \quad g\in G
$$
are required to satisfy equation (4).  Using the fact that $J^{kn}_\lambda =
\lambda^k$, we
obtain
$$
{\partial \over {\partial\ovy}} t_k + {\partial \over {\partial z}}
t_{k-n}= 0\ ,
\leqno(11a)
$$
$$
-{\partial \over {\partial\ovz}} t_k + {\partial \over {\partial y}}
t_{k-n}= 0\ ,
\leqno(11b)
$$
for
$$
k > n \qquad \hbox{or}\qquad k < 0\ .
$$
Equations (11a) and (11b) can be viewed as differential recurrence
relations.
For each choice of a parametrization $\phi$ of Drinfeld-Sokolov flows subject
to
(11a) and (11b) one gets a class of solutions to the SDYM equations.  By
cross differentiation of (11a), (11b) we get that all $t_k{\rm 's}$, $k
\not=
0\ {\rm mod}n$, satisfy the Laplace equation:
$$
{{\partial^2 t_k} \over {\partial y\partial \ovy}} + {{\partial^2 t_k}
\over
{\partial z\partial \ovz}} = 0 \leqno(12)
$$
It is in fact quite easy to solve equations (11a), (11b) in the class of
analytic
functions in $(y, \ovy, z, \ovz)$.  We have
\bigskip

\Theorem  { Given $2(n-1)$ functions $t_k = g_k$,\ $-(n-1) \le k \le -1$ or
$1
\le k \le n-1$, \underbar{analytic} \underbar {in} $(y, \ovy, z, \ovz)$ and
satisfying (12), the general solution to (11a), (11b) is
$$
t_{k+\ell n} = \DD^\ell g_k + \sum^{
\ell - 1} _{m=0} \DD^m \psi _m\ ,\quad 1 \le k \le n-1 \leqno(13a)
$$
$$
t_{k-\ell n} = \EE^\ell g_k + \sum ^{\ell - 1} _{m=0} \EE^m \overline\psi
_m \ ,
\qquad -(n-1) \le k \le -1\ , \leqno(13b)
$$
for $\ell \ge 1$.  By definition
$$
(\DD f )(y, \ovy, z, \ovz) \equiv \int^{\ovz}_0 (\partial_yf)(y, \ovy,
z,\ovz^1)
d\,\ovz ^1 - \int^{\ovy}_0 \partial_zf(y, \ovy^1, z, 0)d\,\ovy ^1 \leqno
(13c)
$$
$$(\EE f)(y,\ovy, z, \ovz ) = \int^y_0 \partial _\ovz f(y^1 , \ovy, z,
\ovz)dy^1 - \int^z_0 \partial _\ovy f(0, \ovy, z^1, \ovz)dz^1\ .\leqno(13d)
$$ The functions $\psi_m (\overline \psi _m)$ are analytic functions of $y,
z(\ovy, \ovz)$, otherwise arbitrary.
}
\bigskip

\noindent
{\bf Examples}\quad Let $t_1 = \alpha y + \beta \ovy + \gamma z + \delta
\ovz +
t^0_1$, \quad $t_k = 0\quad  2 \le k \le n - 1$\ .  Then
$
t_{1+n} = \alpha \ovz - \gamma \ovy + \psi_1(y, z)\ ,
$\quad
$t_{1+2n} = \ovz \partial _y \psi_1 - \ovy \partial _z \psi_1 + \psi_2(y,
z)$ and
so forth.  We can terminate this sequence by choosing $\psi_1 = \psi_2 =
0$\
.   As a special case we might take $n = 2$,\quad $t_1 = \ovy + y$\ , \quad
$t_3 = \ovz$\ ,  which leads to a symmetry reduction of the SDYM to the
(potential) KdV equation discussed in [6], see also [7].  Another
simple choice of $\phi$ obtained when on takes $t_1 = \ovy + y$ \ , \quad $t_3
=
\ovz$\ ,
\quad $t_{-1} = y$\ gives in the case of the maximal parabolic $\tilde \pi =
\{\alpha _1\}$  a solution of the SDYM  which is simultaneously parametrized by
 the (potential)
KdV
equation and the stationary case of the Bogoyavlenskii equation [5] . Such
solutions will in general depend on three independent variables thus giving a
three
dimensional solution of the SDYM.
  For $\tilde \pi =
\emptyset$,
one obtains the modified KdV and the Sine-Gordon equation respectively.
The
choice of $t_1 = \ovy$\ ,\quad $t_{-1} = y$ gives also rise to the stationary
case
of the
Bogoyavlenskii equation.

The above theorem allows one to choose the "times" $t_k$ to have
nonlinear dependence on
$(y, \ovy, z, \ovz)$. We have seen above an example of $\phi$ whose rank is 3.
It is equally easy to give an example of $\phi$ with maximal rank of 4.  As an
example one can take
$$
\eqalign{
t_{-1}& = y^2 + t^0_{-1}\cr
t_1 & = \ovy\, \ovz + zy + t^0_1\cr
t_{n+1} &= z\ovz - y\ovy + t^0_{n+1}\cr
t_{2n+1} &= -\ovy\, \ovz + t^0_{2n+1}
}
$$
all other times being zero.
Thus by employing more then two flows from the soliton hierarchy one can in
principle construct four dimensional solutions of SDYM . It remains an
interesiting
open question as to what class of solutions one obtains this way.

\noindent
{\bf Remark}
One can develop a similar theory to the one presented in this letter starting
with the AKNS systems and the AKNS hierarchy.
It will be published elsewhere.
\bigskip

\noindent
{\bf 4. More Examples}
\itemitem\item{(i)}  rational solutions to the KdV;
\itemitem\item{(ii)} the Bogoyovlenskii equation
 \medskip

It seems that the most interesting cases to consider are classes of
solutions
to generalized Drinfeld-Sokolov hierarchy which are in some sense
finite
dimensional.  The class of rational solutions, soliton solutions, or even
more
generally finite gap solutions are good examples of those. To illustrate
this
point let us consider the rational solutions to the KdV equation vanishing
at
infinity.  By a result of Segal and Wilson [8; section 7] they are
obtained
by factorizing
$$
g_d(\lambda)=\pmatrix { \lambda^{-{{(d+1)}\over 2}} & 0 \cr
0 & \lambda^{ {d+1} \over 2}
} , \qquad d \in \ints_+,  \ d \ \hbox{is odd}
$$
or
$$
g_d(\lambda)=\pmatrix {\lambda^{{d \over 2}}
   & 0 \cr
0 & \lambda^{ -{d \over 2}}} , \qquad d \in \ints_+,  \ d \ \hbox{is even}
$$

The manifold of rational solutions $\MM$ is a union of disconnected pieces
$\MM _ d$ each labeled by a positive integer $d = \dim \MM_d$.  As a
result,
whenever one performs the factorization of
$$
e^{\sum t_k J^k_\lambda} g_d(\lambda)e^{-\sum t_k J^k_{\lambda}}\ ,
$$
$\phi_-$ will depend only on the first $d$ flows $t_1, t_3, \cdots
t_{2d-1}$\
.  We can however easily choose $t_1, t_3, \cdots
t_{2d-1}$ to have, say,
polynomial
dependence on $(y, \ovy, z, \ovz)$ subject to (13a).  Thus we will obtain a
large class of (complex) rational solutions to the SDYM equations labeled
by
$d$. It would be interesting to contrast these with instanton solutions.

To get some insight into the type of solutions we are getting we will consider
two simplest cases corresponding to $d=1$   and $d=2$.
In the first case

$$
g_1(\lambda)=\pmatrix {\lambda^{-1}
   & 0\cr
0 & \lambda},
$$

and the factorization (8) can easily be performed.  The matrix $\phi_-$ depends
now on only one variable, namely $t_1$.  A straightforward computation gives

$$
\phi_-=\pmatrix{1 & -{{1}\over{t_1}}\lambda ^{-1} \cr
0& 1}.
$$
Let us make a choice of parametrization for $t_1, \cdots$ satisfying (11a) and
(11b).  For convenience we set $t_1(y,\ovy,z,\ovz)=\chi$.
It is quite easy to compute now the corresponding self-dual connection.
To this end we use the formulas (6a) and (6b), and $\phi_-$ given above.
The chain rule implies that
$$
A_y=\pmatrix{0 & 0\cr
      -{{\partial \chi}\over{\partial{\ovz}}} & 0},\leqno(15a)
$$
$$
A_{\ovy}=\pmatrix{-{{{\partial \chi}\over{\partial{\ovy}}}\over{\chi}} &
{{{\partial \chi \over {\partial z}} +{{\partial
\chi}\over{\partial{\ovy}}}\chi}
\over {\chi ^2}}\cr
0 &{{{\partial \chi}\over{\partial{\ovy}}}\over{\chi}}},\leqno(15b)
$$
$$
A_z=\pmatrix{0 & 0\cr
      {{\partial \chi}\over{\partial{\ovy}}}& 0},\leqno(15c)
$$

$$
A_{\ovz}=\pmatrix{-{{{\partial \chi}\over{\partial{\ovz}}}\over{\chi}} &
{{-{\partial \chi \over {\partial y}} +{{\partial
\chi}\over{\partial{\ovz}}}\chi}
\over {\chi ^2}}\cr
0 &{{{\partial \chi}\over{\partial{\ovz}}}\over{\chi}}}.\leqno(15d)
$$
One can directly check that indeed we have obtained a self-dual connection.
We would like to point out that (15a)-(15d) give the self-dual connections for
any $\chi$
satisfying the Laplace equation (12).  Moreover, as the next example shows, the
resulting
connections depend in general on all 4 variables $y, \ovy,z,\ovz$ and thus the
method proposed above is capable of producing some 4 dimensional solutions to
the SDYM equations.  Indeed, let us choose
$$
\eqalign{
t_1=\chi & = \ovy\, \ovz + zy, \cr
t_{3} &= z\ovz - y\ovy, \cr
t_{5} &= -\ovy\, \ovz . }
$$
The formulas (15a)-(15d) simplify now to
$$
A_y=\pmatrix{0 & 0\cr
      -\ovy & 0},
$$
$$
A_{\ovy}=\pmatrix{-{{\ovz}\over{yz+\ovy \ovz}} &{{y+\ovz ^3\ovy ^2+2\ovz ^2\ovy
zy+\ovz z^2y^2}\over{(yz+\ovy \ovz)^2}}\cr
0 & {{\ovz}\over{yz+\ovy \ovz}}},
$$
$$
A_z=\pmatrix{0 & 0\cr
      \ovz & 0},
$$
$$
A_{\ovz}=\pmatrix{-{{\ovy}\over{yz+\ovy \ovz}} &{{-z+\ovy ^3\ovz ^2+2\ovy
^2\ovz
zy+\ovy z^2y^2}\over{(yz+\ovy \ovz)^2}}\cr
0 & {{\ovy}\over{yz+\ovy \ovz}}}.
$$
In the  case of $d=2$
$$
g_2(\lambda)=\pmatrix {\lambda
   & 0\cr
0 & \lambda^{-1}},
$$
$$
\phi_-=\pmatrix{1+3{{t_1}\over{(t_1^3-3t_3)}}\lambda ^{-1} &
-3{{t_1^2}\over{(t_1^3-3t_3)}}\lambda ^{-1} \cr
3{{1}\over{(t_1^3-3t_3)}}\lambda ^{-1} & 1-3{{t_1}\over{(t_1^3-3t_3)}}\lambda
^{-1}}.
$$
We now set  $t_1(y,\ovy,z,\ovz)=\chi _1 $ and $t_3(y,\ovy,z,\ovz)=\chi _3$,
both assumed to satisfy (11a) and (11b), and proceed as above.
The final result is [9] :
$$
A_y=\pmatrix{0 & 0\cr
      -{{\partial \chi _1}\over{\partial{\ovz}}} & 0},
$$
$$
A_{\ovy}=\pmatrix{-3{{(3\chi _3+2\chi _1^3){{\partial \chi _1} \over {\partial
z}} -3\chi _1{{\partial \chi _3} \over {\partial z}}+(\chi _1^5 -3\chi _1^2\chi
_3){{\partial \chi _1} \over {\partial \ovy}}}\over{(-\chi _1^3 +3\chi _3)^2}}&
{{(18\chi _1 \chi _3+3\chi_1 ^4){{\partial \chi _1} \over {\partial z}}-9\chi
_1^2{{\partial \chi _3} \over {\partial z}} +(9\chi _3^2-6\chi _1^3\chi _3+\chi
_1^6){{\partial \chi _1} \over {\partial \ovy}}}\over{(-\chi _1^3 +3\chi
_3)^2}}\cr
-3{{3\chi _1^2{{\partial \chi _1}\over{\partial z}}-3{{\partial \chi _3} \over
{\partial z}}+(2\chi _1^4-6\chi_1\chi _3){{\partial \chi _1} \over {\partial
\ovy}}}\over{(-\chi _1^3 +3\chi _3)^2}}
&3{{(3\chi _3+2\chi _1^3){{\partial \chi _1} \over {\partial z}} -3\chi
_1{{\partial \chi _3} \over {\partial z}}+(\chi _1^5 -3\chi _1^2\chi
_3){{\partial \chi _1} \over {\partial \ovy}}}\over{(-\chi _1^3 +3\chi
_3)^2}}},
$$
$$
A_z=\pmatrix{0 & 0\cr
      {{\partial \chi _1}\over{\partial{\ovy}}} & 0},
$$
$$
A_{\ovz}=\pmatrix{-3{{(-3\chi _3-2\chi _1^3){{\partial \chi _1} \over {\partial
y}} +3\chi _1{{\partial \chi _3} \over {\partial y}}+(\chi _1^5 -3\chi _1^2\chi
_3){{\partial \chi _1} \over {\partial \ovz}}}\over{(-\chi _1^3 +3\chi _3)^2}}&
{{(-18\chi _1 \chi _3-3\chi_1 ^4){{\partial \chi _1} \over {\partial y}}+9\chi
_1^2{{\partial \chi _3} \over {\partial y}} +(9\chi _3^2-6\chi _1^3\chi _3+\chi
_1^6){{\partial \chi _1} \over {\partial \ovz}}}\over{(-\chi _1^3 +3\chi
_3)^2}}\cr
-3{{-3\chi _1^2{{\partial \chi _1}\over{\partial y}}+3{{\partial \chi _3} \over
{\partial y}}+(2\chi _1^4-6\chi_1\chi _3){{\partial \chi _1} \over {\partial
\ovz}}}\over{(-\chi _1^3 +3\chi _3)^2}}
&3{{(-3\chi _3-2\chi _1^3){{\partial \chi _1} \over {\partial y}} +3\chi
_1{{\partial \chi _3} \over {\partial y}}+(\chi _1^5 -3\chi _1^2\chi
_3){{\partial \chi _1} \over {\partial \ovz}}}\over{(-\chi _1^3 +3\chi
_3)^2}}}.
$$
Thus we have found this time a self-dual connection parametrized by two scalar
functions $\chi _1$ and
$\chi _2$, which need only satisfy the Laplace equation (12) and the
conditions (11a) and (11b).

Another interesting subject would be to consider systems of equations defined
by the factorization
$$
e^{\sum t_kJ^k_\lambda}  g(w_1; \lambda)e^{-\sum t_kJ^k_\lambda} =
\phi^{-1}_-
(\underline t) \phi_+ (\underline t)\ , \qquad \partial _1 g = \partial _2
g =
0\ .\leqno(16)
$$
The whole theory developed in this paper applies to this factorization without
any major changes.

Since not  much is known about these systems (see [10] for some information) we
restrict  ourselves
to a particularly interesting example which is the (full) Bogoyavlenskii
equation that
corresponds to choosing $n = 2$ and
$$
t_1 = \ovy,\qquad t_{-1} = y\ .
$$
For the maximal parabolic case and $n = 2$ we obtain [5]
$$
u_{\ovy\,\ovz} = u_{\ovy\,\ovy\,\ovy\,y} - 4(u_{\ovy\,\ovy} +
u_{\ovy\,\ovy} u_y +
2u_{\ovy} u_{\ovy y})\ . \leqno(17)
$$
To derive this equation one needs to compute the zero curvature condition
for $U_1$ and $U_{-1}$ using (10a) and (10b).

The proof that this equation is indeed a symmetry reduction of the SDYM is
in [10].

Finally, we would like to point out that the method proposed in this paper
relies heavily
on the factorization problem (5).  The examples of solutions to the SDYM we
have given were
obtained by
 explicitly performing  such a factorization.  However this is not the only
way.
The other way would have been to apply the representation theoretic method [11]
in which the factorizations like the one we are using are recovered from the
appropriate $\tau$ function.
\bigskip\bigskip\bigskip
\centerline{\bf Acknowledgement}\bigskip

I would like to thank S. Dorfmeister and D. Lerner for many stimulating
discussions and   G. Patrick  for his help with Maple.
 I also would like to thank S.
Chakravarty for his interest in my work and for providing me with the
reference [10].  Finally I would like to thank
Prof. M.
Ablowitz for bringing the subject of SDYM reductions to my attention.

\bigskip\bigskip\bigskip

\centerline{\bf References}
\smallskip

\item{1. } Ward, R.S., On self-dual guage fields, Phys. Lett. 61A (1977),
81-82.
\medskip

\item{2. } Dorfmeister, J., Gradl, H. and Szmigielski, J., Group splittings

and equations of Drinfeld-Sokolov type, in preparation.
\medskip

\item{3. }  Drinfel'd, V.G.,and Sokolov,V.V.,
Lie algebras and equations of Korteweg-deVries type,
Journal of Soviet Mathematics, 30, (1985), 1975-2036
\medskip

\item{4. } Kac, V., Infinite dimensional Lie algebras, Cambridge University
Press, Cambridge, U.K., 1985.
\medskip

\item{5. } Bogoyavlenskii, O.I., Breaking solitons in 2+1 - dimensional
integrable equations, Russian Math, Surveys 45:4 (1990), 1-86.
 \medskip

\item{6. } Mason, J. and Sparling, G., Nonlinear Schr\"odinger and Korteweg
-de Vries equations are reductions of self-dual Yang-Mills equations, Phys.

Lett., 137A, 29-33 (1989)
 \medskip

\item{7. } Chakravarty, S. and Ablowitz, M.J., On reduction of self-dual
Yang-Mills equations, in "Painleve Transcendents, their Asymtotics and
Physical Applications". Levi, D. and Winternitz, P. (Eds.), Plenum (1992).
 \medskip

\item{8. } Segal, G. and Wilson, G., Loop groups and equations of KdV type,
Publications I.H.E.S. 61, (1985).
 \medskip
\item{9.} private Maple program g2ex.ml on request from
szmigiel@vincent.usask.ca .
\medskip
\item{10. } Schiff, J., Integrability of Chern-Simons Higgs Vortex Equations
and a reduction of the Self-Dual Yang-Mills Equations to three dimensions, in
"Painleve Transcendents, their Asymtotics and
Physical Applications". Levi, D. and Winternitz, P. (Eds.), Plenum (1992).
 \medskip
\item{11.}Date, E, Jimbo, M., Kashiwara, M., and Miwa, T., Transformation
groups for soliton
equations, in "Nonlinear inegrable systems-classical and quantum theory".
Jimbo, M.,  and Miwa, T.
(Eds.), World Scientific (1983).

\end